\documentclass[12pt]{article}
\def \bea{\begin{eqnarray}}
\def \beq{\begin{equation}}

\def \bo{B^0}

\def \eea{\end{eqnarray}}
\def \eeq{\end{equation}}

\def \ob{\overline{B}^0}

\def \od{\overline{D}^0}

\begin{document}

\begin{flushright}
TECHNION-PH-2006-02\\
EFI 06-02 \\
hep-ph/0601136 \\
January 2006 \\
\end{flushright}

\bigskip
\medskip
\begin{center}
\large
{\bf Isospin in $B$ Decays and the} \\
{\bf $(B^0 \overline{B}^0)/(B^+ B^-)$ Production Ratio
\footnote{To be published as a Brief Report in Physical Review D.}}

\bigskip
\medskip

\normalsize
{\it Michael Gronau and Yuval Grossman} \\

\medskip
{\it Department of Physics, Technion-Israel Institute of Technology \\
Technion City, 32000 Haifa, Israel}

\bigskip
and
\bigskip

{\it Jonathan L. Rosner} \\
\medskip

{\it Enrico Fermi Institute and Department of Physics \\
University of Chicago, Chicago, Illinois 60637} \\

\bigskip
\bigskip

{\bf ABSTRACT}

\end{center}

\begin{quote}

Methods are proposed for measuring the ratio $R^{+/0} = \Gamma(\Upsilon(4S)
\to B^+ B^-)/$ $\Gamma(\Upsilon(4S) \to \bo \ob)$ without assuming isospin
invariance in exclusive hadronic decays such as $B \to J/\psi K$.  The validity
of isospin invariance in various $B$ decays is discussed. Isospin violations
are expected to be especially small in inclusive semileptonic
decays, so that $\Gamma(B^+ \to X_c^0 \ell^+ \nu_\ell) = \Gamma(B^0 \to X_c^-
\ell^+ \nu_\ell)$ is expected to hold at the sub-percent level.  Suggestions
are made for utilizing this relation.  The accuracy of a ``double-tag'' method
for obtaining $R^{+/0}$, such as that used for measuring the $D^+ D^-/D^0 \od$
ratio at the $\psi'' \equiv \psi(3770)$, is also estimated.

\end{quote}

The ratio 
\beq
R^{+/0} = \frac{\Gamma(\Upsilon(4S) \to B^+ B^-)}{\Gamma(\Upsilon(4S) \to\bo
\ob)}
\eeq
is a fundamental quantity in the comparison of $B^+$ and $B^0$ branching
ratios.  In the limit of negligible isospin-violating effects, one expects
$R^{+/0} = 1$, but various Coulomb and other isospin-violating corrections can
induce corrections of as much as 20\% to this value \cite{Voloshin:2003gm}.

The BaBar Collaboration has recently obtained $R^{+/0} = 1.006 \pm 0.036 \pm
0.031$ \cite{Aubert:2004ur}, by assuming equal rates for $B^+ \to J/\psi K^+$
and $\bo \to J/\psi K^0$.  This value dominates the current world average
of $1.020 \pm 0.034$ \cite{Group(HFAG):2005rb}.  The assumption of equal rates
in $B^{+,0}\to J/\psi K^{+,0}$ is based on isospin invariance and on neglecting
$\Delta I=1$ contributions. The latter assumption is reasonable because the
dominant quark process contributing to these decays is $\bar b \to \bar c c
\bar s$, which has $\Delta I=0$.  Effects which result in deviation from equal
rates were estimated in Ref.~\cite{Fleischer:2001cw} to amount to no more than
a percent in rate difference.  However, no rigorous upper bound was
established. Moreover, measuring $R^{+/0}$ by assuming $\Gamma(B^+ \to
J/\psi K^+)=\Gamma(\bo \to J/\psi K^0)$ is not formally correct. It
attempts to measure isospin breaking in one process by assuming
isospin invariance in another.  One would rather measure $R^{+/0}-1$,
which is leading order in isospin breaking, without assuming isospin
symmetry at all, or by assuming it to hold in a process where isospin
breaking is much suppressed.

The BaBar Collaboration also has measured the fraction $f_{00}$ of
$\Upsilon(4S)$ decays to $B^0 \ob$ using a comparison of single and double
production of $D^{*\pm} \ell \nu_\ell$, obtaining $f_{00} = 0.487 \pm 0.010 \pm
0.008$ without assuming isospin invariance \cite{Aubert:2005bq}.  However,
without a direct measurement of the corresponding
ratio $f_{+-}$ of $\Upsilon(4S)$ decays to $B^+ B^-$, one cannot exclude the
possibility of some non-$B \bar B$ decays of the $\Upsilon(4S)$ (for which
only an upper limit of 4\% exists \cite{Barish:1995cx}) which could
lead to differences between $f_{00}$ and $f_{+-}$ of up to a few percent.

In the present note we discuss two alternative methods for determining
$R^{+/0}$ which are not subject to the assumption of isospin in $B \to
J/\psi K$.  The first utilizes
the expectation of equal semileptonic decay rates $\Gamma(B^+ \to X^0_c \ell^+
\nu_\ell) = \Gamma(B^0 \to X^-_c \ell^+ \nu_\ell)$ up to terms of order
$1/m_b^3$.  The second utilizes the ``double tag'' method which has
been successful in measuring the corresponding ratio for $D$ mesons at the
$\psi'' \equiv \psi(3770)$ \cite{Adler:1987as,He:2005bs}.

We begin with a discussion of the order of isospin-violating effects in
exclusive and inclusive $B$ decays.  In general, consider
exclusive final states $f^{+,0}$. If isospin invariance
implies $\Gamma(B^+ \to f^+) = \Gamma(B^0 \to f^0)$, as in the case $f =
J/\psi K$, then 
\beq
A_f \equiv \frac{\Gamma(B^+ \to f^+) - \Gamma(B^0 \to f^0)}
                {\Gamma(B^+ \to f^+) + \Gamma(B^0 \to f^0)}
 = {\cal O}\left( \epsilon_I \right)~~~,
\eeq
where $\epsilon_I\sim (m_d-m_u)/ \Lambda_{QCD}$ with $\Lambda_{QCD}$ a
typical hadronic scale, represents an isospin breaking effect. That
is, the isospin breaking effects enter at leading order in
$\epsilon_I$ and at zeroth order in $1/m_b$
\cite{Bauer:2002aj,Beneke:2000wa}. Basically, while some
isospin-breaking effects do scale like $1/m_b$, there are others, such
as those in form factors, that do not, and are present even in the
$m_b\to\infty$ limit.

On the other hand, for inclusive semileptonic
final states, $f = X_c \ell^+ \nu_\ell$, one has
\beq
A_{X_c \ell \nu} = {\cal O}\left( \frac{\epsilon_I} {m_b^2} \right)~~.
\eeq
In inclusive semileptonic decays to charm, one can use heavy quark
symmetry to classify the terms which affect semileptonic decay rates 
\cite{Manohar:2000dt}.  At leading order, the decay is described by a decay of
a free $b$ quark. There are no $O(1/m_b)$ terms and at order $1/m_b^2$ there
are two: a kinetic term, parameterized by $\lambda_1$, and a QCD hyperfine
term, parameterized by $\lambda_2$.  These terms depend on the isospin of the
spectator quark.  We generally expect the isospin breaking in $\lambda_i$
($i=1,2$) to be of order $\epsilon_I$. For example, 
\beq
\lambda_2^{(u)}-\lambda_2^{(d)}={1 \over 4}
\left[m^2(B^{+*})-m^2(B^+)-m^2(B^{0*})+m^2(B^0)\right].
\eeq
Thus, the total effect of isospin breaking on inclusive decays is
parametrically much smaller than that expected in $R^{+/0}-1$. That
is, it is very small, of order, $\epsilon_I \lambda_i/m_b^2 <
10^{-3}$ and can be safely neglected.

Isospin-breaking effects for exclusive semileptonic decays were
studied in
\cite{Voloshin:1998nh}. It was shown that linear isospin breaking
terms in $B \to D$ semileptonic decay distributions are zeroth order
in $1/m_b$; however, they become ${\cal O}(\epsilon_I^2)$ when
integrated over phase space. Thus, also for exclusive semileptonic
decays, isospin-breaking effects can be neglected compared to those
expected in $R^{+/0}-1$.

We now discuss the semileptonic branching ratios to charmed final states 
for $B^+$ and $B^0$, denoted by ${\cal B}_{+,c}$ and ${\cal B}_{0,c}$,
which are expected if $\Gamma(B^+ \to X_c^0 \ell^+ \nu_\ell) = 
\Gamma(B^0 \to
X_c^- \ell^+ \nu_\ell)$.  The $B$ semileptonic branching ratio, averaged over
$B^+$ and $B^0$ decays, is $\overline{\cal B}_{\rm SL} = (10.95 \pm 0.15)\%$
\cite{Group(HFAG):2005rb}.  From this one must subtract approximately 2\% of
its value, or 0.22\%, for $b \to u$ semileptonic decays, leaving $(10.73 \pm
0.15)\%$.  We take account of the lifetime ratio $r_\tau \equiv \tau(B^+)/\tau
(B^0) = 1.076 \pm 0.008$ \cite{Group(HFAG):2005rb}. In order to obtain separate
branching ratios for $B^+$ and $B^0$ one may assume that $\overline{\cal
B}_{\rm SL}$ is due to equal $B^+$ and $B^0$ contributions. The effect of 
a difference between these two contributions on determining
${\cal B}_{+,c}-{\cal B}_{0,c}$ is second order in isospin breaking, involving
the product $(R^{+/0}-1)(r_{\tau}-1)$.  One then expects
$$
{\cal B}_{+,c} \equiv
{\cal B}(B^+ \to X_c^0 \ell^+ \nu_\ell) = (11.12 \pm 0.16)\%~~,~~~
$$
\beq \label{eqn:brcpreds}
{\cal B}_{0,c} \equiv
{\cal B}(B^0 \to X_c^- \ell^+ \nu_\ell) = (10.34 \pm 0.15)\%~~.
\eeq

It is easier to measure the total semileptonic rate, including the small
contribution of $X_u$ final states.  In principle the so-called ``weak
annihilation'' contribution \cite{Bigi:1993bh} can lead to differences between
$\Gamma(B^+ \to X_u^0 \ell^+ \nu_\ell)$ and $\Gamma(B^0 \to X_u^- \ell^+
\nu_\ell)$.  Upper limits on this process have been placed
recently~\cite{Rosner:2006zz}.  These effects vanish in the $m_b\to \infty$
limit and will be neglected.  In that case one expects
$$
{\cal B}_+ \equiv
{\cal B}(B^+ \to X^0 \ell^+ \nu_\ell) = (11.35 \pm 0.16)\%~~,~~~
$$
\beq \label{eqn:brpreds}
{\cal B}_0 \equiv
{\cal B}(B^0 \to X^- \ell^+ \nu_\ell) = (10.55 \pm 0.15)\%~~.
\eeq
and it is these predictions (and particularly their ratio ${\cal B}_+/{\cal
 B}_0 = r_\tau$) that one wishes to test.

Consider a total sample of $N_+$ charged $B$ and $N_0$ neutral $B$ mesons.
(At present both BaBar and Belle have accumulated several hundred million $B
\bar B$ pairs.)  Inclusive $B$ semileptonic decays are to be identified
opposite fully reconstructed $B^-$ or $\ob$ mesons.  In what follows, sums
of the quoted states and their charge conjugates are implied unless otherwise
noted.

Take the example of charged $B$ mesons; similar considerations apply to neutral
$B$'s.  Assume an efficiency $\epsilon_-$ for fully reconstructing a produced
$B^-$, and an efficiency $\epsilon^{\rm SL}_+$ for detecting a $B^+$
semileptonic decay.  The total number of tagged $B^+$ opposite a reconstructed
$B^-$ will be $N_+^{\rm tag} = \epsilon_- N_+$, while the number of detected
semileptonic decays will be $\epsilon_- N_+ \epsilon^{\rm SL}_+ {\cal B}_+$.
The relative statistical error on this quantity will be the reciprocal of its
square root.  Taking representative numbers of $N_+ = 2.5 \times 10^8$,
$\epsilon_- = 5 \times 10^{-3}$ \cite{Aubert:2005sb}, $\epsilon^{\rm SL}_+ =
0.5$, and ${\cal B}_+ = 11.35\%$, one finds
\begin{equation} \label{eqn:brerr}
\frac{\Delta {\cal B}_+}{{\cal B}_+} = \left( \frac{2.5 \times 10^8}{N_+}
\right)^{1/2} \left( \frac{5 \times 10^{-3}}{\epsilon_-} \right)^{1/2} \left(
\frac{0.5}{\epsilon^{\rm SL}_+} \right)^{1/2} \left( \frac{11.35\%}
{{\cal B}_+} \right)^{1/2} \times 0.38\%~~.
\end{equation}
A similar calculation may be applied to $\Delta {\cal B}_0/{\cal B}_0$; the
nominal reconstruction efficiency for $B^0$'s at BaBar is $\epsilon_0 = 3
\times 10^{-3}$ \cite{Aubert:2005sb}.
The dominant errors in the determination of ${\cal B}_{+,0}$ and their ratio
are unlikely to be statistical.  They are likely to arise from uncertainties in
the efficiencies $\epsilon^{\rm SL}_{+,0}$ and hence will require Monte Carlo
simulation.  In any case there seem to be no problem to attain a statistical
accuracy on ${\cal B}_+/{\cal B}_0$ at the percent level with present BaBar
and Belle samples.

Note that in order to measure the branching ratios one does not need to know
the tagging efficiencies $\epsilon_{-,0}$, at least if these do not depend on
whether a semileptonic decay is observed opposite the tagged meson.  However,
in order to learn the ratio $R^{+/0} = N_+/N_0$ one needs the tagging
efficiencies as well as the lifetime ratio $r_\tau = \tau_+/\tau_0$.
In that case one has
\beq
{N^{SL}_+ \over N^{SL}_0} =
{\epsilon_- \epsilon^{SL}_+ \over \epsilon_0 \epsilon^{SL}_0}
 R^{+/0}  {\tau_+ \over \tau_0}
\eeq
The accuracy of this determination is likely to be governed almost completely
by the accuracy of prediction or measurement of the tagging efficiencies.

In the absence of sufficiently precise information on $\epsilon_{-,0}$
one may attempt to measure the ratio of semileptonic $B^+$ and $B^0$ decays
{\it inclusively} at the $\Upsilon(4S)$.  This would require identifying
whether a given lepton comes from a vertex with an even ($B^0$) or odd
($B^+$) number of tracks.  While resolutions on individual tracks at Belle
and BaBar are unlikely to be good enough to permit this, one might be able
to use events with two opposite-sign primary leptons from $B$ and $\bar B$
decays to help define two disjoint vertices, particularly if one is willing
to select events with long $B$ and $\bar B$ lifetimes \cite{ARPC}.

Since isospin breaking is also expected to be small in exclusive semileptonic
modes \cite{Voloshin:1998nh}, one could obtain $R^{+/0} = f_{+-}/f_{00}$ by
measuring $f_{+-}$, since $f_{00}$ is already known to a couple of percent of
its value \cite{Aubert:2005bq}.  Comparison of single and double $D^* \ell
\nu_\ell$ production in the same manner as was used to obtain $f_{00}$ is
difficult because the soft-charged-pion signature of $D^{*\pm} \to \pi^\pm D^0
(\od)$ which was so useful in measuring $f_{00}$ via $D^{*\pm} \ell \nu_\ell$
production is not available in $D^{*0}$ decay.  One might be able to make use
of the soft {\it neutral} pion in $D^{*0} \to D^0 \pi^0$ to measure $f_{+-}$
via single and double exclusive semileptonic decay.
This method was used in Ref.\ \cite{Athar:2002} to obtain $f_{+-}/f_{00}
= 1.058 \pm 0.084 \pm 0.136$ based on CLEO II data.

We now evaluate the expected accuracy of a double-tag method for determining
$R^{+/0}$.  Again we concentrate on determining the number of charged $B$'s;
similar methods apply to neutral $B$'s.  The number of singly-tagged $B^+$
is $N_+^{\rm single} = N_+ \epsilon_-$, while the number of
doubly-reconstructed $B^+ B^-$ pairs is $N_{\rm chg}^{\rm dbl} = N_+ \epsilon_-
\epsilon_+^{(-)}$.  Here $\epsilon_+^{(-)}$ is the number of $B^+$ which can
be reconstructed {\it opposite an already-reconstructed $B^-$}.  We expect
that $\epsilon_+^{(-)} \ge \epsilon_-$.  In other words, once a $B^-$ has
already been reconstructed it should be at least as easy to reconstruct its
$B^+$ partner as it would have been to reconstruct a $B^-$ or $B^+$ alone.

The efficiencies roughly cancel out in the expression
\beq
N_+ = \frac{(N_+^{\rm single})^2}{N_{\rm chg}^{\rm dbl}}
 \frac{\epsilon_+^{(-)}}{\epsilon_-}~~~.
\eeq
A Monte Carlo simulation is needed to estimate $\epsilon_+^{(-)}/\epsilon_-$.
The relative statistical error on $N_+$ is dominated by that on
$N_{\rm chg}^{\rm dbl}$ and is
\beq
\left| \frac{\Delta N_+}{N_+} \right| \simeq (N_{\rm chg}^{\rm dbl})^{-1/2}~~.
\eeq
For a 1\% determination of $N_+$ one needs $N_{\rm chg}^{\rm dbl} \simeq 10^4$,
which would require $4 \times 10^8$ $B^+$ events if both $\epsilon_-$ and
$\epsilon_+^{(-)}$ were $5 \times 10^{-3}$.  The improvement of reconstruction
efficiencies beyond this value thus is of prime importance.

To conclude, we have discussed alternative methods for measuring the isospin 
breaking quantity, $R^{+/0}-1$ which are based on semileptonic decays.  The
theoretical advantage of these methods over methods based on exclusive hadronic
decays such as $B\to J/\psi K$ is that isospin breaking in semileptonic decays
is known to be parametrically smaller than in $R^{+/0}-1$.

\medskip
We thank Romulus Godang, Zoltan Ligeti, Dan Pirjol, Aaron Roodman and 
Marie-Helene Schune for useful discussions.  J. L. R.
wishes to thank the Technion -- Israel Institute of Technology for gracious
hospitality during part of this work.
This work was supported in part by the United States Department of
Energy through Grant No.\ DE FG02 90ER40560, by the Israel Science 
Foundation under Grants No. 1052/04 and 378/05, and by the German--Israeli 
Foundation under Grant No. I-781-55.14/2003.


\begin{thebibliography}{99}

\bibitem{Voloshin:2003gm}
  M.~B.~Voloshin,
  Mod.\ Phys.\ Lett.\ A {\bf 18}, 1783 (2003)
  [arXiv:hep-ph/0301076], and earlier references therein;
  R. Kaiser {\it et al.}, Phys.\ Rev.\ Lett.\ {\bf 90}, 142001 (2003);
  Soo-hyeon Nam, Mod.\ Phys.\ Lett.\ A {\bf 20}, 95 (2005).

\bibitem{Aubert:2004ur}

B.~Aubert {\it et al.}  [BABAR Collaboration],
  Phys.\ Rev.\ D {\bf 69}, 071101 (2004)
  [arXiv:hep-ex/0401028]; and earlier references therein.

\bibitem{Group(HFAG):2005rb}
  Heavy Flavor Averaging Group(HFAG),
  ``Averages of $b$-hadron properties as of winter 2005,''
  arXiv:hep-ex/0505100.  For updated version, prepared after the summer and
fall 2005 conferences, see {\tt http://www.slac.stanford.edu/xorg/hfag/}.

\bibitem{Fleischer:2001cw}
  R.~Fleischer and T.~Mannel,
  Phys.\ Lett.\ B {\bf 506}, 311 (2001)
  [arXiv:hep-ph/0101276].

\bibitem{Aubert:2005bq}
  B.~Aubert {\it et al.}  [BABAR Collaboration],
  Phys.\ Rev.\ Lett.\  {\bf 95}, 042001 (2005)
  [arXiv:hep-ex/0504001].

\bibitem{Barish:1995cx}
  B.~Barish {\it et al.}  [CLEO Collaboration],
  Phys.\ Rev.\ Lett.\  {\bf 76}, 1570 (1996).

\bibitem{Adler:1987as}
  J.~Adler {\it et al.}  [MARK-III Collaboration],
  Phys.\ Rev.\ Lett.\  {\bf 60}, 89 (1988).

\bibitem{He:2005bs}
  Q.~He {\it et al.}  [CLEO Collaboration],
  Phys.\ Rev.\ Lett.\  {\bf 95}, 121801 (2005)
  [arXiv:hep-ex/0504003].

\bibitem{Bauer:2002aj}
  C.~W.~Bauer, D.~Pirjol and I.~W.~Stewart,
  Phys.\ Rev.\ D {\bf 67}, 071502 (2003)
  [arXiv:hep-ph/0211069].

\bibitem{Beneke:2000wa}
  M.~Beneke and T.~Feldmann,
  Nucl.\ Phys.\ B {\bf 592}, 3 (2001)
  [arXiv:hep-ph/0008255].

\bibitem{Manohar:2000dt}
For a review see, for example,
  A.~V.~Manohar and M.~B.~Wise,
  Camb.\ Monogr.\ Part.\ Phys.\ Nucl.\ Phys.\ Cosmol.\  {\bf 10}, 1 (2000).

\bibitem{Voloshin:1998nh}
  M. B. Voloshin,
  Phys.\ Lett.\ B {\bf 433}, 419 (1998)
  [arXiv:hep-ph/9803243].

\bibitem{Bigi:1993bh}
  I. I. Y. Bigi and N. G. Uraltsev,
  Nucl.\ Phys.\ B {\bf 423}, 33 (1994)
  [arXiv:hep-ph/9310285].

\bibitem{Rosner:2006zz}
  J.~L.~Rosner {\it et al.} [CLEO Collaboration],
  arXiv:hep-ex/0601027;
  T. O. Meyer,
  ``Limits on weak annihilation in inclusive charmless semileptonic B decays,''
Ph.\ D. Thesis, Cornell University, 2005, UMI-31-62893,\\
{\tt http://www.slac.stanford.edu/spires/find/hep/www?r=umi-31-62893}.

\bibitem{Aubert:2005sb}
  B.~Aubert {\it et al.}  [BABAR Collaboration],
  presented at EPS International Europhysics Conference on High Energy Physics
  (HEPP-EPS 2005), Lisbon, Portugal, 21--27 July 2005, arXiv:hep-ex/0507085.

\bibitem{ARPC} We thank Aaron Roodman for this suggestion.

\bibitem{Athar:2002}
 S. B. Athar {\it et al.} [CLEO Collaboration],
  Phys.\ Rev.\ D {\bf 66}, 052003 (2002).

\end{thebibliography}
\end{document}